\documentclass[
reprint,
 aps,
prb,
]{revtex4-1}

\usepackage{graphicx}% Include files
\usepackage{dcolumn}% Align table columns on decimal point
\usepackage{bm}% bold math
\usepackage{amsmath}
\begin{document}

\preprint{APS/123-QED}

\title{Large deviation theory to model systems under an external
feedback}

%\collaboration{}%\noaffiliation

\author{Alessio Gagliardi$^1$, Alessandro Pecchia$^{2}$, Aldo Di Carlo$^3$}

\affiliation{ (1) Technische Universitaet Muenchen, Arcisstrasse 21, 80333, Munich (Germany) \\
alessio.gagliardi@tum.de. \\
(2) CNR, Via Salaria Km 29,600, 0017 Monte Rotondo (Italy) \\
(3) University of Rome "Tor Vergata", Via del Politecnico 1,
00133, Rome (Italy)}%

\date{\today}% It is always \today, today,

\begin{abstract}
In this paper we address the problem of systems under an external
feedback. This is performed using a large deviation approach and
rate distortion from information theory. In particular we define a
lower boundary for the maximum entropy reduction that can be
obtained using a feedback apparatus with a well defined accuracy
in terms of measurement of the state of the system. The large
deviation approach allows also to define a new set of potentials,
including information, which similarly to more conventional
thermodynamic potentials can define the state with optimal use of
the information given the accuracy of the feedback apparatus.
\begin{description}
\item[PACS numbers]
\end{description}
\end{abstract}

\maketitle

%\tableofcontents

\section{Introduction}

The idea of a thermodynamic theory of systems under an external
feedback dates back to the origin of statistical physics when
Maxwell made a gedanken experiment about the work that could be
extracted by a system controlled by an external apparatus. In
principle the external feedback can probe the state of the system
and manipulates it in order to extract work. In his first
formulation the feedback was treated like an oracle that can make
measures and manipulate the system without any cost in terms of
work and entropy production, for this reason it was called a
"Demon" to stress its abstract nature. However the idea was
reconsidered along all the past century \cite{dem1,dem2} with
different approaches and conclusions. One of the central work on
the topic was the paper by Szilard and his famous Szilard engine
\cite{szilard}, a still very idealized machine that anyway shows
already some more realistic characteristics in terms of its
components and the nature of the feedback controller. Starting
from Szilard engine, but generalizing the concept, several authors
\cite{info1,info2,info3,info4,info5,info6,info7} and in particular
Sagawa and Ueda, have developed a new branch of thermodynamics of
systems under an external feedback \cite{sagawa1,sagawa2,sagawa3},
i.e. information thermodynamics.

In particular in \cite{sagawa5} it was shown, using an ingenious
extension of fluctuation theorems including the information
gathered by the feedback, that the presence of the external
controller can increase the average work that can be extracted
from a system when it is driven from one equilibrium state to
another according to
\begin{equation}\label{int1}
    \langle W \rangle \leq \Delta F + k_{B} T I,
\end{equation}
with $\langle W \rangle$ the average work, $\Delta F$ free energy
variation between the final and initial equilibrium states, $k_B$
is the Boltzmann constant, $T$ the bath temperature and $I$ the
mutual information. For a cyclic transformation, for which $\Delta
F = 0$ the maximum work reduces to $k_{B} T I$.

The mutual information, for a classical system, is described by
the mutual information functional of information theory. There is
an understandable debate about the validity of using information
theoretical quantities and their interpretations in thermodynamics
\cite{hanggi}. However, for many cases of interest \cite{sagawa4}
Shannon entropy represents a good choice for the entropy
functional, especially in equilibrium cases. In particular Shannon
entropy gives the right entropy functional but only physical and
experimental evidences can determine which is the correct
probability density function (PDF) of the problem under
investigation, as information theory has nothing to say about that
\cite{alessio1}.

An interesting alternative approach to non equilibrium
thermodynamics is arising within the framework of a different
branch of information theory, e.g. large deviation theory (LDT).
LDT is a mathematical framework used to establish the probability
of fluctuations of statistical quantities from their {\em typical
values}. Typicality in information theory leads to the definition
of thermodynamic averages and state variables
\cite{touchette1,merhav2}.

In particular, LDT demonstrates that for a broad class of PDFs,
the probability of fluctuations from the average value drops
exponentially with the fluctuation magnitude. The exponent is
proportional to the number of degrees of freedom of the system,
explaining why, in the thermodynamic limit, fluctuations of state
variables become negligible. However in the emerging field of
stochastic thermodynamics which also copes with small systems in
non equilibrium conditions, fluctuations can be an important
aspect of their behavior \cite{exp1,exp2,exp3,exp31,exp33}.

Many systems of interest have the peculiarity of being the
interconnection between a physical system probed and controlled by
a feedback apparatus, for example biological processes in cells
and other living organisms belong to this class. Recently, several
studies have investigated the effect of including information
terms in the analysis of biological processes within the framework
of information thermodynamics with great success, see for example
\cite{barato,Ito,exp4,exp5,exp6,exp7,exp8,exp9}. Thus the
investigation of thermodynamics under feedback is rapidly rising
interest in many fields beyond statistical physics.

In this paper we give a new insight to the problem of a system
under an external feedback using a special case of LDT and the
concept of typicality. The link between systems under feedback and
LDT is obtained using rate distortion theory (RDT), a fundamental
part of information theory. The chain of relationships between LDT
and RDT that we outline, is particularly pleasing as it makes a
direct connection between the information the feedback controller
apparatus gathers about a system and the associated entropy
reduction, also leading to an explicit construction of a
thermodynamic potential for a system under feedback. The
possibility to construct potentials including the effect of
information opens the perspective of a complete new analysis of
those systems more similar to conventional equilibrium
thermodynamic formalism. For a similar approach see also
\cite{merhav23}.

The paper is organized as follows: in the first part a brief
summary of the main information theoretical concepts, Shannon
entropy, conditional entropy and mutual information is given. Then
the concept of typicality is introduced. In the next section we
present the large deviation theory and the code large distortion
problem. The final part is devoted to introducing rate distortion
theory and the final link to thermodynamics under an external
feedback.

\section{A brief excursus in Information theory}

The central quantity of information theory is the Shannon entropy
defined as
\begin{equation}\label{inf1}
    S(P) = -k_B \sum_{i} p_i \ln p_i,
\end{equation}
where $p_i$ represents the probability of the $i^{th}$ event and
$P$ the PDF. Shannon entropy is usually expressed in bits and
adimensional, we have chosen here the convention to include the
Boltzmann constant and express the entropy using natural
logarithm. However, this is totally immaterial for the present
discussion. Shannon entropy plays a central role in
thermodynamics, even for a large class of systems under non
equilibrium conditions \cite{sagawa4}.

More generally if we have a PDF $p(\xi)$ which depends on a set of
degrees of freedom, $\xi$, we can define the proper discrete (or
continuous) Shannon entropy. The degrees of freedom depend on the
problem at hand, they could be the set of positions and momenta of
a collection of particle in gas phase for example. From the
Shannon entropy it is possible to promptly derive other four
connected quantities. The first is the joint (Shannon) entropy
which is just the entropy of two random vectors $\xi$ and $\pi$:
\begin{equation}\label{inf2}
    S(\Xi, \Pi) = -k_B \sum p(\xi, \pi) \ln p(\xi, \pi),
\end{equation}
with $p(\xi, \pi)$ the joint probability. If the two vectors are
independent the joint entropy reduces to the sum of the individual
entropies, in the other case we have:
\begin{equation}\label{inf3}
    S(\Xi, \Pi) = S(\Xi) + S(\Pi) - k_B I(\Xi, \Pi),
\end{equation}
with $I(\Xi, \Pi)$ the mutual information functional defined as:
\begin{equation}\label{inf4}
    I(\Xi, \Pi) = \sum p(\xi, \pi) \ln \left [ \frac{p(\xi, \pi)}{p(\xi)p(\pi)} \right
    ].
\end{equation}
The mutual information is a special case of the Kullback-Leibler
divergence (KLd) defined as:
\begin{equation}\label{inf5}
    D(P \| Q) = \sum  p(\xi) \ln \left ( \frac{p(\xi)}{q(\xi)} \right ),
\end{equation}
where $P$ and $Q$ are two PDFs. The KLd is a pseudo distance
between PDFs and has an important role in LDT due to the Chernoff
bound \cite{merhav2}, in stochastic thermodynamics \cite{seifert},
within fluctuation theorems \cite{ft3} and in entropy production
within the Boltzmann equation \cite{villani}. In particular, we
observe that the mutual information has the form of a KLd between
the joint PDF, $p(\xi, \pi)$, and the independent PDF, $p(\xi,
\pi)=p(\xi) p(\pi) \equiv q(\xi, \pi)$.

Finally, mutual information and entropy of two random vectors are
connected by the conditional entropy:
\begin{equation}\label{inf6}
    S(\Xi|\Pi) = S(\Xi) - k_B I(\Xi, \Pi) = \sum p(\xi,
    \pi)\ln p(\xi|\pi).
\end{equation}
Conditional entropy represents the entropy (uncertainty) left in
$\Xi$ after the conditioning to $\Pi$.

These five are the most relevant functionals in information theory
as practically every important theorem is related to one or
another in some form.

\section{Typical set in the phase space}

\begin{figure}[htb]
\begin{center}
\includegraphics*[width=8cm, angle=0]{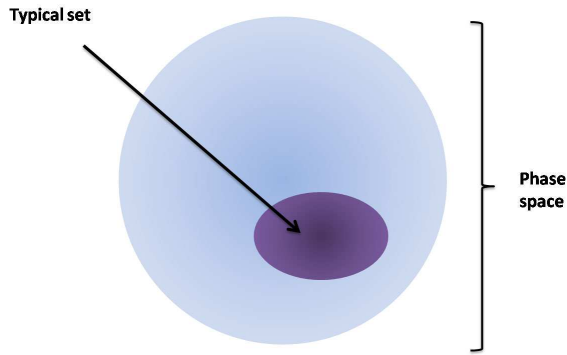}
\caption{\it{\small{(color online) The phase space and the typical
subset. Usually, except for the uniform distribution, the typical
set is indeed a proper subset of the entire class of possible
microstates in the phase space. Its volume grows exponentially
with the entropy of the problem.}}} \label{fig:typ}
\end{center}
\end{figure}

The Shannon entropy has a nice geometrical interpretation in the
typical set theorem consequence of the asymptotic equipartition
principle \cite{cover}. The theorem states that given a PDF
$p(\xi)$, where $\xi$ is the vector of degrees of freedom of the
problem, with entropy $S(\Xi)$ then the "typical set" within the
phase space has a volume equal to:
\begin{equation}\label{typ1}
    \Omega_{typ} \sim e^{\frac{S(\Xi)}{k_B}},
\end{equation}
where the meaning of "typical" means that the probability for the
system to be found in a microstate within the typical set
converges to 1 in the thermodynamic limit, namely,
\begin{equation}\label{typ2}
    p(\xi \, \in \, \Omega_{typ}) \rightarrow 1.
\end{equation}
In other words the concept of typicality states that only a
portion of the entire phase space is really relevant to compute
ensemble averages of thermodynamic quantities, considering that
the volume $\Omega_{typ}$ fundamentally collects all the
probability (see Fig.~\ref{fig:typ}).

In particular for any quantity, $A(\xi)$, defined over the phase
space, the ensamble average is defined as,
\begin{equation}\label{typ3}
   \langle A \rangle = \int \tilde{A} q(\tilde{A}) d\tilde{A},
\end{equation}
with
\begin{equation}\label{typ4}
    q(\tilde{A}) = \int p(\xi) \delta(\tilde{A} -
    A(\xi) ) d\xi.
\end{equation}
If it happens that for all microstates within the typical set,
$A(\xi \, \in \,\Omega_{typ}) = A^{*}$ (constant), then $A$ is a
state variable of the problem and $\langle A \rangle$ = $A^{*}$.

Several generalizations of the typical set concept exist for
example for a joint PDF in the joint typical set theorem
\cite{cover}. The concept of typicality is extremely important
also in thermodynamics. In the microcanonical ensemble where the
PDF over the phase space is a uniform PDF, the entropy reduces to
the integration of the density of accessible microstates.

However, as recent studies on stochastic thermodynamics have shown
\cite{hanggi,seifert}, it is possible to extend many thermodynamic
results also for small systems, i.e., systems where significant
fluctuations of the state variables are not only possible, but
also probable. Such type of systems are also those for which a
practical implementation of a feedback controller is more feasible
due to their limited number of degrees of freedom.

\section{Large deviation theory}

There is an elegant formalism within information theory connecting
the statistical analysis of microscopic fluctuations with the
macroscopic behavior of the system described by thermodynamic
potentials, this formalism is large deviation theory (LDT). This
connection is a well established fact dating back to the '70,
several authors \cite{touchette2,ellis,ellis2,ellis3,lanford,oono}
used LDT to derive many results of equilibrium thermodynamics. In
particular it was possible to derive in a very elegant way the
maximum entropy principle/ minimum free energy for systems in the
thermodynamic limit in microcanonical or canonical ensemble.

The entire idea of LDT is to estimate the probability of
fluctuations departing from the average in stochastic problems.
This can be directly applied to evaluate the probability of
observing a fluctuation of a state variable in a thermodynamic
system.

Let us assume we have a quantity $A$ with average value $A^{*}$
and $N$ degrees of freedom in the system. We define the
contribution to $A$ per degrees of freedom $a$ = $A/N$ and $a^{*}$
= $A^{*}/N$. It is said that a stochastic problem follows a Large
deviation (LD) law if the probability that $a$ departs from
$a^{*}$ follows an exponential law:
\begin{equation}\label{ldt1}
    q( a \neq a^{*}) \sim e^{-N K(a)},
\end{equation}
where $K(a)$ is an exponent which is dependent on the
thermodynamic quantity, while $q(a)$ is defined as in
eq.~\ref{typ4} from the PDF per microstate $p(\xi)$.

A powerful theorem to check if a problem satisfies a LD law is the
G$\ddot{a}$rtner-Ellis theorem (GET) \cite{ellis}. We first define
the Scaled Cumulant Generating Function (SCGF) as
\begin{equation}
\label{ldt2}
   \lambda(\alpha) = \lim_{N \to \infty} \frac{1}{N} \ln \left [ \langle
e^{N\alpha a}\rangle \right ],
\end{equation}
with
\begin{equation}\label{ldt3}
    \left \langle e^{N\alpha a} \right \rangle = \int p(\xi) e^{\alpha N
    a(\xi)} d\xi,
\end{equation}
with $\alpha$ a real number. The GET states that if the SCGF
exists and is differentiable everywhere in $\alpha$, then the
system fluctuations follow an exponential law.

We notice that the SCGF is very similar to a partition function,
but scaled with respect to $N$ and also where every exponential
term is weighted by the probability per microstate, $p(\xi)$. If
we make a simple variable change $\alpha$ = $-\beta$, the new
parameter $\beta$ plays the same role as the inverse temperature,
$\beta$ = $1/(k_B T)$ and the SCGF can be rewritten as
$-\phi(\beta)$ = $\lambda(\alpha)$. The GET does not only provide
a condition for existence, but also gives an operative way to
evaluate the exponent $K(a)$. In fact it demonstrates that $K(a)$
is related to the Legendre-Fenchel transform of the SCGF:
\begin{equation}
K(a)  =  - \min_{\beta \, \geq \, 0} [\beta a - \phi(\beta)].
\label{ldt4}
\end{equation}

It is possible to demonstrate \cite{merhav3} that the previous
equation can be rewritten in terms of the real partition function
(without the PDF weighting the exponents as in the SCGF), but
adding a constant:
\begin{equation}
    J(a)  =  - \min_{\beta \, \geq \, 0} [\beta a - \phi(\beta)]
     + \frac{1}{N}\ln \Lambda = K(a) + \frac{1}{N}\ln \Lambda.
\label{ldt5}
\end{equation}
The latter constant has the form of the entropy of a uniform PDF
$U(\xi) = 1/\Lambda$, with $\Lambda$ a particular volume within
the phase space (see \cite{merhav3}), which depends on the prior
PDF $p(\xi)$. In the case of a microcanonical ensemble it reduces
to all the microstates with the same energy $\bar{E}$. The
function $\phi(\beta)$ is related to the free energy potential.
Specifically, we have that the free energy per degrees of freedom
($f = F/N$) is equal to:
\begin{equation}\label{ldt6}
    f = \frac{\phi(\beta)}{\beta},
\end{equation}
thus the Legendre-Fenchel transform of the SCGF is linked to the
entropy of the system per degree of freedom, $s(a)$, by
\begin{equation}
     J(a)  =  - \frac{s(a)}{k_B} + \frac{1}{N}\ln \Lambda.
\label{ldt7}
\end{equation}

If we consider the constant term as the entropy associated to a
uniform PDF we have that the exponent $J(a)$ can be treated as
follows:
\begin{eqnarray}
    J(a) & = & \frac{- s(a)}{k_B} + \frac{1}{N}\ln \Lambda
    \nonumber \\
    & = &  \frac{1}{N} \left ( \sum_{\xi} p(\xi) \ln p(\xi)  + \sum_{\xi}
    p(\xi) \ln \Lambda \right ) \nonumber \\
    & = & \frac{1}{N} D(\Xi // U),
\label{ldt8}
\end{eqnarray}
where $U = 1/\Lambda$ is the uniform PDF over the phase space
volume $\Lambda$, $D$ is the KLd and $Ns(a) = S_{tot} = -\sum
p(\xi) \ln p(\xi)$ the total entropy. Thus the probability $q(a
\neq a^{*})$ goes like the following:
\begin{equation}\label{ldt9}
    q(a \neq a^{*}) \approx e^{-D(\Xi // U)},
\end{equation}
recovering the Chernoff bound for large fluctuations within the
typical set formalism \cite{cover}. The LDT provides a clear
understanding why entropy maximization is at the essence of
equilibrium thermodynamics. Similar results are obtained for a
canonical ensemble \cite{touchette1}. For a general discussion
about the GET and the LDT applied to thermodynamics we refer to
\cite{ellis}.

Notably, LDT has been applied to non equilibrium systems
\cite{touchette2} substituting the PDF for a microstate with the
PDF of entire time dependent trajectories to take into account the
time evolution of the system. Even more important, LDT can be used
to derive fluctuation theorems. In fact if the exponent has the
symmetry relation, e.g. $K(-a) - K(a) = \gamma a$ ($\gamma$ is a
positive real number) for a certain thermodynamic quantity, $A =
Na$ , we immediately get the fluctuation theorem
\cite{touchette1,seifert,FT1a,FT2a,FT3a}:
\begin{equation}\label{ldt10}
    \frac{q(a)}{q(-a)} \approx e^{N(K(-a) -
    K(a))}= e^{N\gamma a}.
\end{equation}

A dual theorem of GET is the Varadhan theorem \cite{varhadan}
which allows to invert the Legendre-Fenchel transform. This states
that if
\begin{equation}
    K(a)  =  - \min_{\beta \, \geq \, 0} [\beta a - \phi(\beta)],
\label{ldt11}
\end{equation}
is valid, then also the following relation holds:
\begin{equation}
    \phi(\beta)  = \min_{a} [\beta a + K(a)].
\label{ldt12}
\end{equation}
The LD law in microcanonical or canonical ensemble explains,
through the Varadhan theorem, why the minima of the free energy
potentials are associated to the state variable values for the
equilibrium state.

\section{Rate distortion theory and large deviation theory}

\begin{figure}[htb]
\begin{center}
\includegraphics*[width=8cm, angle=0]{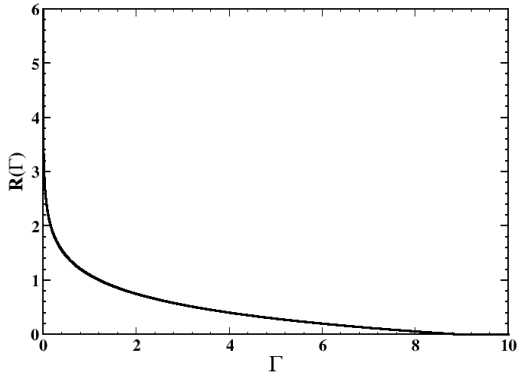}
\caption{\it{\small{Typical shape of a RDF. In this case it is
plot the RDF of a Gaussian PDF with variance $\sigma^{2} = 9$. For
$\Gamma$ larger than $\sigma^{2}$ the RDF is 0.}}} \label{fig:rdf}
\end{center}
\end{figure}

In this paragraph we make explicit the connection between LDT and
the information theory quantities presented in the previous
sections and consider a system under feedback. In order to do so
we need to introduce the rate distortion function (RDF), a central
functional in rate distortion theory (RDT). RDT copes with a
fundamental problem in communication, namely estimating the
minimal information content that any message sent over a
communication channel must contain such that the receiver can
still have a good reconstruction of the original signal. The error
source can be either due to distorting noise, a finite channel
capacity or because of a lossy compression operated by the sender.
In mathematical form, if we define a distance, $d(\pi,\xi)$,
between the message sent, $\xi$, and the message received, $\pi$
(eventually after decompression, noise deconvolution, etc...), the
target of RDT is to find under which conditions the average
distance can be kept lower than a given threshold, $\Gamma$
\cite{cover}. Formally we ask,
\begin{equation}\label{rdt1}
    \langle d(\xi, \pi) \rangle \leq \Gamma,
\end{equation}
where the average is computed over the joint PDF $p(\xi, \pi)$.
The distance, $d$, can be any functional with the properties of a
distance (symmetry, positive definite, $d=0 \, \, iif \, \,
\xi=\pi$, and must satisfy the Schwartz inequality). The most
important theorem of RDT states that, given $d$ and $\Gamma$,
there exists a function, $R(\Gamma)$  (the rate distortion
function), representing the minimum information required in order
to send messages with an average distortion not greater than
$\Gamma$. The function $R(\Gamma)$ has some remarkable properties:
it is convex in the argument $\Gamma$, it converges to the entropy
of the source (for a discrete PDF) for $\Gamma = 0$, while for
$\Gamma \geq \Gamma^{*}$ it is zero. In practice $\Gamma^{*}$ is
the limit distortion value after which the information content is
completely lost and the receiver has equal chance by just guessing
at random the most likely message, based on the joint PDF
\cite{cover}. For an example of a RDF for a Gaussian PDF see
Fig.~\ref{fig:rdf}.

The second important aspect of the rate distortion function is
that it can be computed as a constrained minimization of a mutual
information functional,
\begin{equation}
\label{rdt2}
   R(\Gamma) = \min_{p(\pi|\xi): \,  \langle d(\xi, \pi) \rangle \leq \Gamma}
   I(\Xi, \Pi).
\end{equation}
In the latter the minimization is with respect to the conditional
PDF, $p(\pi|\xi)$, and the constrain is that the average
distortion remains always smaller than $\Gamma$.

The connection between LDT and RDT is materialized by the
distortion coding problem (DCP) \cite{merhav}. DCP can be
formulated in the following way: let assume we have two random
vectors $\xi$ and $\pi$ with PDFs $p(\xi)$ and $q(\pi)$,
respectively, and we want to know what is the probability that
picking up at random two vectors $\xi$ and $\pi$, one from each
distribution, the distance is such that $d(\xi,\pi) \leq \Gamma$.
If we assume the only constrain that $\xi$ should belong to the
typical set of $\Xi$, the probability of such condition follows a
LDT with an exponent equal to the rate distortion function:
\begin{equation}\label{rdt3}
    p(\xi, \, \pi: \, \langle d(\xi ,\pi) \rangle \leq \Gamma) \propto
    e^{-R(\Gamma)}.
\end{equation}
The function $R(\Gamma)$ monotonically decreases in the range $0 <
\Gamma < \Gamma^{*}$, with limiting values $R(0)=S/k_b$, $R(\Gamma
> \Gamma^{*}) = 0$.

In the two limiting cases: when $\Gamma = 0$, $R(0) = S(\Xi)/k_B$.
In the case $\Gamma \geq \Gamma^{*}$ the probability converges to
1 because $R(\Gamma \geq \Gamma^{*}) = 0$.

Since the rate distortion function appears in the form of a LD
law, there is a second way to define the rate distortion function
using the GET and the Legendre-Fenchel transform of the SCGF
\cite{berger,merhav}:
\begin{equation}\label{rdt4}
     R(\Gamma) = - \min_{\lambda \geq 0} \left [ \lambda \Gamma  + \sum p(\xi) \ln (Z_{\pi}(\lambda) ) \right ],
\end{equation}
with
\begin{equation}\label{rdt5}
    Z_{\pi}(\lambda) = \sum_{\pi} q(\pi) e^{-\lambda d(\xi,\pi)}.
\end{equation}
$Z_{\pi}$ has the form of a generalized partition function, linked
to the distribution of $\pi$, where the distance $d$ has a role
similar to energy in the conventional partition function
\cite{gray}.

An interesting aspect of this particular case of the LDT is that
minimizing $R(\Gamma)$ w.r.t. $\lambda$, we get an important
relation,
\begin{equation}\label{rdt6}
    \Gamma = \sum_{\xi , \pi} d(\xi,\pi) \frac{p(\xi) q(\pi) e^{-\lambda^{*}
    d(\xi,\pi)}}{Z_{\pi}(\lambda^{*})} = \langle d(\xi,\pi) \rangle.
\end{equation}
The average is made with respect to the joint PDF,
\begin{equation}\label{rdt7}
    \tilde{p}(\xi,\pi) = \frac{p(\xi) q(\pi) e^{-\lambda^{*}
    d(\xi,\pi)}}{Z_{\pi}(\lambda^{*})},
\end{equation}
which is the joint probability that fulfill the minimal rate
function for a defined average error $\Gamma$ and $\lambda^{*}$ is
the value minimizing the Legendre-Fenchel transform.

\section{Large deviation, rate distortion and feedback control}

We can now use the results of RDT applied to LDT to obtain our
most important result concerning the thermodynamics of systems
with feedback control. Let assume that we have a feedback
controller that performs measurements of the state of a system and
then afterwards manipulates it. If we assume that the measurement,
$\pi_k$, has some correlation with the state $\xi$, then the
entropy of the system after measurement will be given by the PDF,
$p(\xi| \pi = \pi_k)$, conditioned by the outcome $\pi_k$. This is
equal to:
\begin{equation}\label{fa1}
    S(\Xi| \pi = \pi_k) = -k_B \sum p(\xi| \pi = \pi_k) \ln p(\xi| \pi =
    \pi_k).
\end{equation}
A typical set volume is always associated to this entropy,
\begin{equation}\label{fa2}
    \Omega_{typ}(\xi| \pi = \pi_k) = e^{\frac{S(\Xi| \pi =
    \pi_k)}{k_B}},
\end{equation}
hence the average volume of the typical set after a measurement is
thus:
\begin{equation}\label{fa3}
    \left \langle \Omega_{typ}(\xi| \pi = \pi_k) \right \rangle
    = \left \langle e^{\frac{S(\Xi| \pi =
    \pi_k)}{k_B}} \right \rangle,
\end{equation}
where the average is made w.r.t. $q(\pi)$.

The lower boundary of this formula can be obtained using the
Jensen inequality, thanks to the convexity of the exponential
function, $\langle exp(f) \rangle \geq exp(\langle f \rangle)$, to
obtain:
\begin{equation}\label{fa4}
    \left \langle \Omega(\xi| \pi = \pi_k)_{typ} \right \rangle
    \geq e^{ \frac{\langle S(\Xi| \pi =
    \pi_k) \rangle}{k_B}} = e^{\frac{S(\Xi|\Pi)}{k_B}},
\end{equation}
exploiting the fact that $\langle S(\Xi| \pi = \pi_k) \rangle =
S(\Xi|\Pi)$.

Thus, we find that the feedback can -at best- constrain the volume
of phase-space by the conditional entropy. Splitting the
conditional entropy in the usual, $S(\Xi|\Pi) = S(\Xi) - k_B I(\Xi
, \Pi)$, we obtain that the effect of the feedback is to compress
the volume of the typical set with an exponent at best equal to
the mutual information:
\begin{equation}\label{fa5}
   \Omega_{typ}(\xi|\pi) \geq \Omega_{typ}(\xi) e^{-I(\Xi,\Pi)}.
\end{equation}

The fact that the mutual information is always non negative
ensures that the effect of the feedback is always to compress the
original typical set.

If we want to connect the mutual information to the effective
measurement operated by the feedback apparatus (FA), then we can
define a distance functional, $d$, and an average distortion
$\Gamma$ between state and estimation and use the rate distortion
function to find the minimal mutual information required to have a
certain average distortion $\Gamma$. The final result is the lower
boundary,
\begin{equation}\label{fa6}
    \left \langle \Omega_{typ}(\xi| \pi = \pi_k) \right \rangle
    \geq \Omega_{typ}(\xi) e^{-R(\Gamma)}.
\end{equation}
The appealing of equation (\ref{fa6}) is that it gives a direct
connection between the effect of feedback information and the
effective operative measurement performed by the FA.

A final note regarding the effect of the FA. The reduction in the
volume of the typical set remains only "virtual" until the FA does
not operate and manipulate the system. We can thus imagine this
shrinking of the typical set linked to the $R(\Gamma)$ as the best
average reduction in uncertainty about the state of the system
from the FA side once the measurements have been made. It is
anyway clear that the entropy and the physics of the system are
left unchanged until the FA does not directly operate.

\section{Towards a thermodynamic potential including information}

A thermodynamic potential is nothing else than a quantity that
when minimized/maximized allows to find the equilibrium or
stationary states of a certain thermodynamic system, subject to
given constrains. We have already discussed in section 4 and 5 how
the LDT provides an elegant way to derive the maximal entropy and
minimal energy principles for the microcanonical and canonical
ensembles. In particular, we saw how the free energy potential
comes as part of the Legendre-Fenchel transform of the large
deviation exponent thanks to the Gartner-Ellis theorem. In this
section we put together this result and the results of section 6
in order to construct an explicit thermodynamic potential for a
system under feedback control.

First of all we define a simplified feedback apparatus (SFA). We
assume a system in thermal equilibrium with an external bath at
temperature $T_0$ and we assume that the system can be coupled to
the bath or disconnected from that and coupled to the SFA. The SFA
is a system that can make a measurement and manipulate the system
accordingly in a time $\tau$ much smaller than any relaxation time
of the system. Moreover, we assume a certain level of ideality in
the feedback, i) during the measurement no perturbation of the
system occurs, ii) the feedback uses the entire information during
the manipulation achieving the maximal efficiency. Once the system
has been manipulated by the feedback it is allowed to relax and
finally it is connected again to the external bath until
thermalize with it. This simple cycle assures that every time the
feedback performs its new measurement/manipulation the system is
at thermal equilibrium with temperature $T_0$.

With these approximations we can decouple the initial state before
the feedback operation from the feedback action. A completely
different and more complex scenario occurs in the case when the
system is not allowed to thermalize or the feedback acts
continuously in such a way that the system state depends on
previous feedback history.

We finally assume that the feedback can obtain an estimate, $\pi$,
of the real state, $\xi$,  of the system such that $\langle
d(\xi,\pi) \rangle \leq \Gamma$ for a well-defined distance
functional.

We have shown in section 6 that the feedback action entails to an
entropy reduction after measurement by a factor $I(\xi, \pi) =
R(\Gamma)$. Thus we can finally define the maximum increase in
free energy due to the presence of the feedback:
\begin{eqnarray}
    \Delta F & = & F - F_{eq}  \nonumber \\
    & = & U - T_0 (S - k_B R(\Gamma)) - U +
    T_0 S = k_B T_0 R(\Gamma) = \langle W \rangle. \nonumber \\
    & &
\label{esa}
\end{eqnarray}
with $U$ the internal energy. It recovers the result found by
Sagawa with eq. \ref{int1}, as the maximum work for a cyclic
transformation, but now with a direct connection between the type
and accuracy of the measurement, $d(\xi,\pi)$ and $\Gamma$, and
the average extracted work $\langle W \rangle$.

The $R(\Gamma)$ exponent of equation (\ref{fa6}), following a LD
law, can be expanded in the Legendre-Fenchel transform as in
equation (\ref{ldt4}), leading to a joint probability between the
microstate and the guess equal to equation (\ref{rdt6}). Using the
Varadhan theorem we can invert the Legendre-Fenchel transform for
the RDF and obtain something equivalent to a thermodynamic
potential for the information:
\begin{equation}
     \left \langle \ln (Z_{\pi}(\lambda) \right \rangle =
     \sum p(\xi) \ln (Z_{\pi}(\lambda) ) =   \min_{\Gamma} \left [ \lambda \Gamma  + R(\Gamma) \right ].
\label{rdtI1}
\end{equation}
The structure of this equation is similar to the one of the free
energy in standard thermodynamics with the relation between free
energy and partition function, where now the Chernoff coefficient
$\lambda$ plays the role of inverse temperature:
\begin{equation}\label{rdtI2}
    \phi_{I}(\lambda) \equiv -\frac{1}{\lambda} \left \langle \ln (Z_{\pi}(\lambda) \right
    \rangle.
\end{equation}

What is the meaning of this functional? We can understand its role
by calculating the distance, using the KLd, between a generic
joint probability $p(\xi, \pi)$ and the optimal $\tilde{p}(\xi,
\pi)$ as defined in eq. (\ref{rdt6}):

\begin{eqnarray}
  &  & D(P//\tilde{P}) = \sum_{\xi \pi} p(\xi, \pi) \ln \frac{p(\xi, \pi)}{\tilde{p}(\xi, \pi)} \nonumber \\
   &=&  \sum_{\xi \pi} p(\xi, \pi) \log \frac{p(\xi, \pi)}{p(\xi) q(\pi)} + \lambda^{*}\langle d(\xi, \pi) \rangle +  \langle \ln Z_{\pi}(\lambda^{*}) \rangle \nonumber\\
   &=&  I(\Xi, \Pi) + \lambda^{*}\langle d(\xi, \pi) \rangle -
   \lambda^{*}\phi_{I}(\lambda^{*}).
\end{eqnarray}
The latter equation can be rewritten as:
\begin{equation}\label{rdtI2}
    \lambda^{*}\phi_{I}(\lambda^{*}) = \lambda^{*}\langle d(\xi, \pi) \rangle
    + I(\Xi, \Pi) - D(P//\tilde{P}).
\end{equation}
Considering that the three terms in the rhs are all positive it is
easy to demonstrate that the maximum of the potential
$\phi_{I}(\lambda^{*})$ is obtained for $D(P//\tilde{P}) = 0$,
that is when the joint probability is the ideal one for which the
RDF is achieved.

\section{A simple physical model}

\begin{figure}[htb]
\begin{center}
\includegraphics*[width=8cm, angle=0]{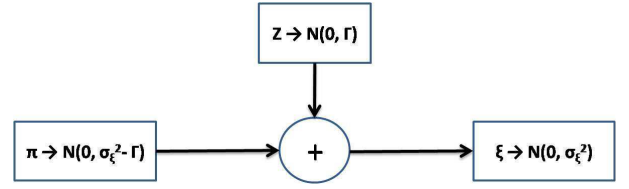}
\caption{\it{\small{(color online) Scheme of a channel between two
joint gaussian random variables.}}} \label{fig:feed}
\end{center}
\end{figure}

We apply the formalism to a simple model: a set of single
particles in a box in gas phase. We assume that every particle is
under a feedback and that it is in thermal equilibrium with a bath
at temperature $T$. The Hamiltonian of the system is given by
$\mathcal{E} = Ap^{2}$, where $A=1/2m$, being $m$ the particle
mass. The particle state, $\xi=(r,p)$, is characterized by an
homogeneous spatial distribution for the position $r$ within the
box, and a normal distribution of the momentum, $p$. Therefore,
neglecting position, we identify the particle state just with the
momentum, $x=p$. The equilibrium distribution is a normal
distribution with 0 mean value and a variance $\sigma^{2}_{\xi}$ =
$k_B T/2A$. The model includes a feedback that can probe the
momentum of the particle. The feedback uses a distance $d(\xi,\pi)
= (\xi - \pi)^{2}$. This measurement is affected by error, so the
measurement $\pi$ is a random variable, statistically correlated
with the dynamical state of the particle. The model assumes that
between every probe and manipulation the system is allowed to
relax to thermal equilibrium, that means that at every measurement
the system is found in the same equilibrium state. Finally, we
also assume that $\pi$ is distributed like a normal random
variable. The situation can be formalized like in \cite{cover} by
assuming that the feedback and the source are connected by a
channel with Gaussian noise (see Fig. \ref{fig:feed}).

The Gaussian noise has distribution:
\begin{equation}\label{mod1}
    p(z) = N(0,\Gamma) = \frac{1}{\sqrt{2 \pi \Gamma}}
    e^{-\frac{z^{2}}{2\Gamma}}.
\end{equation}
For simplicity we assume that the source has mean value equal to
zero ($\mu_{\xi} = 0$), $p(\xi) = N(0, \sigma^{2}_{\xi})$. This
assumption is totally immaterial for the generality of the
discussion. Following \cite{cover} the feedback has a PDF equals
to:
\begin{equation}\label{mod2}
    p(\pi) = N(0, \sigma^{2}_{\xi} - \Gamma) = \frac{1}{\sqrt{2 \pi (\sigma^{2}_{\xi} - \Gamma)}}
    e^{-\frac{\pi^{2}}{2(\sigma^{2}_{\xi} - \Gamma)}}.
\end{equation}
In the latter we have assumed that $0 \leq \Gamma \leq
\sigma^{2}_{\xi}$. With the previous defined distance functional
the average distance $\Gamma$ is equal to the mean square error
$\langle (\xi - \pi)^{2} \rangle$.

For this simple model the form of the rate distortion function is
well known:
\begin{equation}\label{mod3}
    R(\Gamma) = \frac{1}{2} \ln \left (  \frac{\sigma^{2}_{\xi}}{\Gamma} \right
    ),
\end{equation}
with $R(\Gamma) = 0$ for $\Gamma \geq \sigma^{2}_{\xi}$.

It is a simple matter of calculation to demonstrate that indeed
eq.~\ref{rdt4}, using $p(\xi)$, $p(\pi)$ and $d(\xi,\pi) = (\xi -
\pi)^{2}$, gives back the correct rate distortion function. The
relation between $\lambda$ and $\Gamma$ is very simple:
\begin{equation}\label{mod4}
    \lambda = \frac{1}{2\Gamma}.
\end{equation}
We can insert the PDF and the distance functional in
eq.~\ref{rdt7} in order to get the optimal joint distribution for
such marginal PDFs. The solution is a joint Gaussian distribution:
\begin{eqnarray}
    & & \tilde{p}(\xi,\pi) =  \nonumber \\
    & & \, \, \frac{1}{2\pi \sigma_{\xi} \sigma_{\pi} \sqrt{1 - \rho^{2}}}
    exp \left ( -\frac{1}{2(1-\rho^{2})} \left [ \frac{\xi^{2}}{\sigma^{2}_{\xi}}  - \frac{2\rho \xi \pi}{\sigma_{\xi}\sigma_{\pi}} + \frac{\pi^{2}}{\sigma^{2}_{\pi}} \right ]
    \right ) \nonumber \\
    & &
\label{mod5}
\end{eqnarray}
with correlation coefficient equal to $\rho = \sqrt{1 -
\Gamma/\sigma^{2}_{\xi}}$. Using the result in eq.~\ref{mod4} we
obtain the relation between $\lambda$ and the correlation
coefficient:
\begin{equation}\label{mod6}
    \lambda = \frac{1}{2\sigma_{\xi}^{2}(1 - \rho^{2})}.
\end{equation}

As expected if $\Gamma$ is related to the distance between state
and estimation by the feedback, $\lambda$ has a simple
interpretation in terms of correlation between the two random
variables. In case of random vectors (with $N$ elements) the total
average distance $ \Gamma = \sum N \Gamma_{i}$, with $\Gamma_{i}$
the distance associated to the $i^{th}$ component. Clearly the
value $\Gamma$ has an extensive characteristic being function of
the number of degree of freedom ($N$) of the system. On the
contrary $\lambda$ is the intensive counterpart being connected to
the correlation between state and estimation.

In this particular case we can also calculate the optimal work
that such a feedback apparatus can recover from the system as a
function of the average distance $\Gamma$, using eq.\ref{esa}:
\begin{equation}\label{work1}
  \langle W \rangle = Tk_B R(\Gamma) = -\frac{T k_B}{2} \ln ( 1 -
  \rho^{2}) = T k_B I(\Xi, \Pi).
\end{equation}
The latter shows that the maximum work extracted is proportional
as should be to the mutual information between two Gaussian random
variables.

\section{Conclusions}

In this work we have analyzed a system under feedback using
typicality and large deviation theory. In particular this
connection assures to make a simple and natural link between the
way the measurement is performed and its accuracy and the maximum
work that can be extracted in terms of entropy reduction. Clearly
in this discussion we have not considered the effect of the
manipulation from the feedback and its cost in terms of overall
efficiency, thus all our results must be considered as boundary
limits.

The main result of the paper is the possibility, using rate
distortion theory, of developing the equivalent of thermodynamic
potentials even in the case of systems under an external feedback.

This perspective to the problem is particular appealing not only
because it establishes a nice relation between the measurement and
the respective entropy and information functional, but also
because it can be generalized for a large class of problems where
large deviation applies.

\section{Acknowledgements}

We acknowledge Prof. Merhav for the useful comments.

%Equation (\ref{eq1}) can be rewritten as,
%\begin{equation}\label{eta2}
%    \delta S = S[p^{\varepsilon}] - S[p] + 2\varepsilon S[p] - \varepsilon\lambda_{+} -
%    \varepsilon\lambda_{-}.
% + \varepsilon\Pi.
%\end{equation}
%we can finally regroup the entropy perturbation as:
%\begin{eqnarray}
%   \delta S & = & k (1 - 2\varepsilon) D[p_{n}\|p^{\varepsilon}_{n}] + k \varepsilon
%   D[p_{n+1}\|p^{\varepsilon}_{n}] \nonumber \\
%   & &+ k  \varepsilon  D[p_{n-1}\|p^{\varepsilon}_{n}]
%   \nonumber \\
%   & &+ k \varepsilon (p_{\alpha} \ln(p_{\alpha}) + p_{\beta}\ln(p_{\beta}) \nonumber \\
 %  & & - p_{\alpha} \ln(p^{\varepsilon}_{\alpha}) - p_{\beta} \ln(p^{\varepsilon}_{\beta}) \nonumber \\
%   & = & k \varepsilon D[p_{n+1}\|p^{\varepsilon}_{n}] + k \varepsilon D[p_{n-1}\|p^{\varepsilon}_{n}] \nonumber \\
%   & & + k (1 - 2\varepsilon)\rho
%\label{main}
%\end{eqnarray}
%The term $\rho$ in the rhs can be expanded as [REF]:



\end{document}